\documentclass[referee]{raa}

\usepackage[english]{babel}
\usepackage{graphicx}
\usepackage{times}             
\usepackage{amsmath}
\usepackage{amssymb}
\usepackage{natbib}
\bibpunct{(}{)}{;}{a}{}{,}
\usepackage[pagebackref=true]{hyperref}

\usepackage{color}
\usepackage{ulem}
\newcommand{\red}{\textcolor{red}}

\definecolor{mygreen}{rgb}{0.42,0.56,0.14}

\begin{document}
\title{An Approach to Gamma-Hadron Separation via Measurements of the Middle-UV Fraction of EAS emission by an Imaging Atmospheric Cherenkov Telescope 
with UV-Sensitive SiPM detectors} 

   \volnopage{Vol.0 (20xx) No.0, 000--000}      
   \setcounter{page}{1}          

\author{E.E.~Kholupenko\inst{1} 
\and D.V.~Badmaev \inst{1}
\and N.M.~Budnev \inst{2}
\and A.M.~Bykov \inst{1} 
\and A.M.~Krassilchtchikov \inst{1}
\and L.A.~Kuzmichev \inst{2,3} 
\and A.A.~Bogdanov \inst{1} 
\and Yu.V.~Chichagov \inst{1} 
\and G.A.~Repman \inst{1}
\and Yu.V.~Tuboltsev \inst{1}}   

\institute{Ioffe Institute, Saint-Petersburg, Russian Federation; {\it eugene@astro.ioffe.ru} \\
\and Irkutsk State University, ISU, Irkutsk, Russian Federation \\
\and Skobeltsyn Institute of Nuclear Physics, MSU, Moscow, Russian Federation \\
\vs\no
   {\small Received 20xx month day; accepted 20xx month day}}

\abstract
{
The operation of a small-size Cherenkov gamma-ray telescope TAIGA-IACT with camera on SiPMs 
OnSemi MicroFJ-60035 has been modelled by multiparticle Monte Carlo (MC) methods. 
The model implies that telescope camera is equipped with two specific types of 
filters of 290 -- 590 nm (visible+NUV) and 220 -- 320 nm (MUV+UVB)-bands, 
each covering half of the camera pixels in some uniform order. This allows one 
to measure the fraction of UV-radiation in total amount of Cherenkov radiation 
of an extensive air shower (EAS), that can be used for efficient gamma-hadron 
separation. The corresponding quality factor takes values up to 5.07 in the 
10 -- 100 TeV range depending on the distance to EAS axis and camera orientation.
}


   \authorrunning{E.E.~Kholupenko, D.V.~Badmaev, N.M.~Budnev et al.}            
   \titlerunning{Gamma-Hadron Separation by Measurement of EAS MUV-radiation with an IACT on SiPMs}  

   \maketitle

\section{Introduction}
\label{Introduction}
In recent decades Cherenkov gamma-ray astronomy has demonstrated a rapid development (e.g., 
\citealt{Bykov_et_al2017} and references therein): the maximal size of telescope mirrors has significantly 
increased (from 0.9~m (\citealt{Fegan_et_al1968}) to 28~m (\citealt{Gottschall_et_al2015})), the employed photosensors 
are substantially improved (in particular, older photo-multiplier tubes are replaced by newer PMTs with higher 
efficiency (e.g. \citealt{Hanna_et_al2022})), the number of telescopes and detector units operating together 
within a single observatory has increased from a single standalone imaging atmospheric telescope (e.g., Whipple 
(\citealt{Weekes_et_al1989})) to observatories consisting of multiple IACTs and muon detectors (e.g. HEGRA 
(\citealt{Aharonian_et_al1991}), CANGAROO II (\citealt{Tsuchiya_et_al2004}), MAGIC II (\citealt{Tridon_et_al2010}), VERITAS 
(\citealt{Holder_et_al2008}), H.E.S.S. II (\citealt{Hofmann2003}), TAIGA (\citealt{Budnev_et_al2022})). The largest of 
currently developed gamma-ray and cosmic ray observatories will consist of several tens of IACTs (e.g. CTA (\citealt{Longo2022})). 
Such a development has already allowed specialists to achieve very impressive results in observation of cosmic gamma-ray
sources (e.g. \citealt{vanEldik_et_al2015, Ansoldi_et_al2016}). 
These results are based not only on technical improvements, but also on development of data analysis 
techniques: from simple measuring of gamma-ray signal excess above the cosmic ray background 
(e.g. \citealt{Charman_Drever1969}) to analysis of Hillas parameters (\citealt{Hillas1985, Weekes_et_al1989}), 
multifractal and wavelet moments (e.g. \citealt{Razdan_et_al2002} and references therein), 
time- (e.g. \citealt{Aharonian_et_al1997, Razdan_et_al2002}) and spectral parameters (e.g. 
\citealt{Stepanian_et_al1983, Rahman_et_al2001}) and further to analysis involving neural networks (e.g. 
\citealt{Bussino_Mari2001, Razdan_et_al2002, Postnikov_et_al2018} and references therein). 

\begin{figure*}
\centering
\includegraphics[height=10cm,width=14cm]{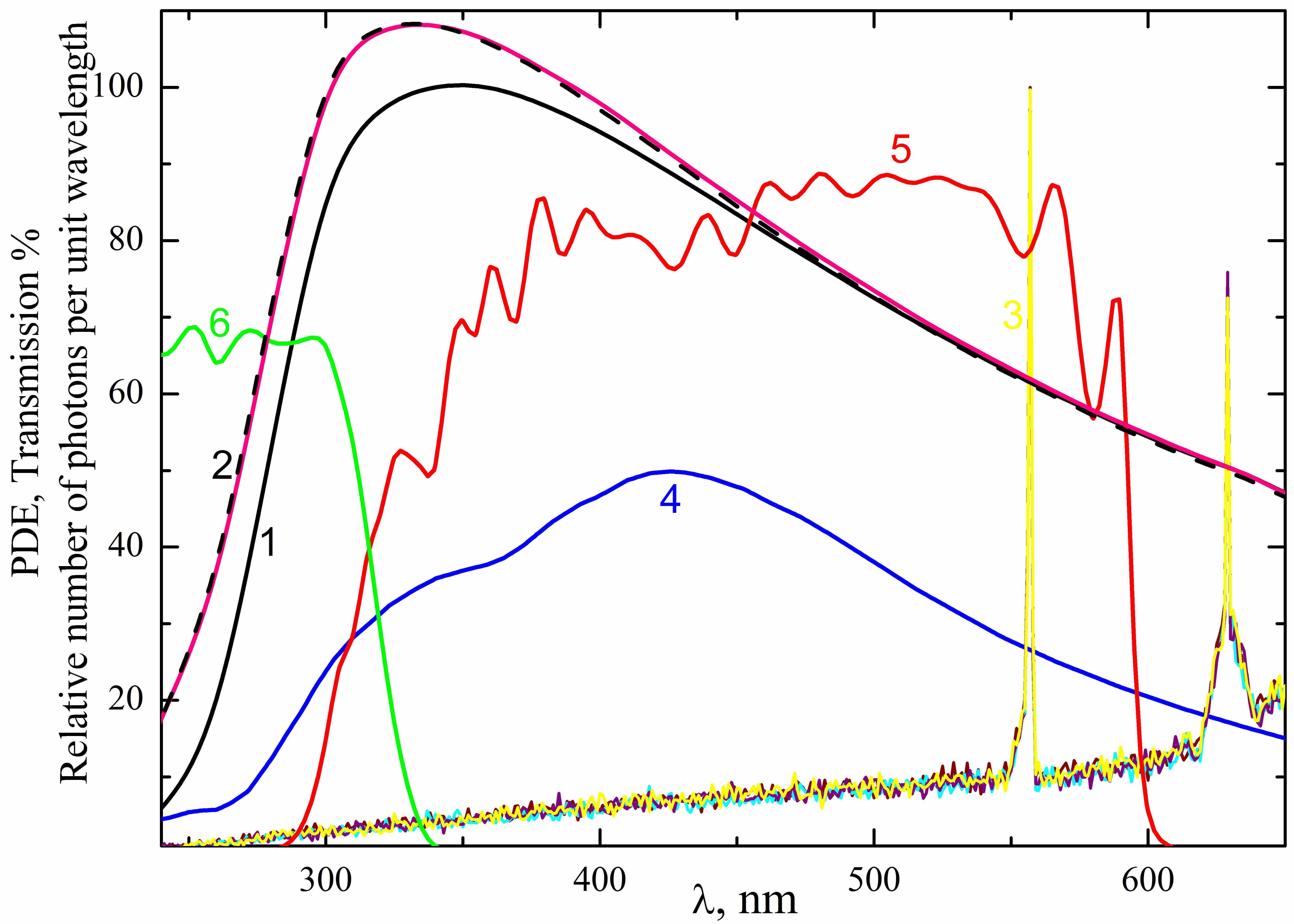}
\caption{
Dependencies of profiles on wavelength $\lambda$:
1) an averaged spectrum of Cherenkov radiation from EASs induced by 31.62 TeV gamma-quanta, 
normalized to 100\% at maximum at $\simeq$340~nm, is shown by the black solid curve; 
2) averaged spectra of Cherenkov radiation from EASs induced by 31.62 TeV and 100 TeV protons, normalized such that to coincide assymptotically with averaged spectrum from $\gamma$-EASs at long wavelengths, 
are shown by pink solid and black dashed curves overlapped practically; 
3) four specific realizations of the night sky background spectrum (normalized to 100\% at maximum at $\simeq$557~nm), simulated within a Monte Carlo approach using the data of 
\cite{Leinert_et_al1998, Benn_Ellison1998, Mikhalev_et_al2001, Mikhalev_Medvedeva2002}, are shown by partially overlapped yellow, purple, cyan, and wine curves; 
4) the photon detection efficiency of a OnSemi MicroFJ-60035 SiPM (\citealt{ON_Semi_Data_Sheet}), PDE$\left(\lambda\right)$, is shown by the blue curve; 
5) the transmission coefficient of the optical band filter SL~290-590 (\citealt{Photooptic_2020}) is shown by the red curve; 
6) the transmission coefficient of the 220-320~nm band NSH+KCSH filter (\citealt{Manomenova_et_al2005, Rudneva_et_al2018}) is shown by the green curve. 
}
\label{fig1}
\end{figure*}

At the same time, some of the developed techniques of observations and data analysis in Cherenkov gamma-ray 
astronomy yet have not become widespread, probably due to relative complexity of realization and application in 
comparison with other competing technologies employed to solve same problems, or to incapability of the existed 
equipment to ensure their effective employment in full degree. 
One of such techniques is measurement of the fraction of ultraviolet (UV) radiation in the total amount 
of extensive air shower (EAS) Cherenkov radiation. This method can be used to perform gamma-hadron separation 
of the shower primaries (hereinafter ``UV gamma-hadron separation''), which is an essential step in data analysis. The approach is based on the fact that 
this fraction for the EASs induced by cosmic ray (CR) protons is larger than for the EASs induced by gamma-quanta, 
because generation of EAS Cherenkov radiation by CR protons in the atmosphere occurs deeper (on average) 
than in the case of EASs from gamma-quanta. Thus, the EAS Cherenkov photons from CR protons propagate 
to the registration point through a smaller optical depth of the atmospheric ozone 
(and, correspondingly, are less absorbed, that is essential in the UV band) in comparison with 
ones from gamma-quanta (for a detailed consideration of this effect, see 
\citealt{Aharonian_et_al1997, Rahman_et_al2001, Razdan_et_al2002, Kholupenko_et_al2018} and references therein).
This, in turn, leads to formation of EAS Cherenkov radiation spectra which have significantly different shapes in UV-range 
(especially, in MUV-range) for EASs induced by gamma-quanta and protons (see Fig. \ref{fig1}). 
For the first time the said technique was proposed by \cite{Stepanian_et_al1983}, tested (\citealt{Zyskin_et_al1987}) 
and employed for registration of cosmic gamma-ray sources (e.g. \citealt{Kalekin_et_al1995, Neshpor_et_al2001}) at 
the GT-48 Cherenkov telescope. The cameras of GT-48 telescopes (as well as other Cherenkov telescopes of the 
first three generations\footnote{Except FACT (\citealt{Anderhub_et_al2011}), 
which is one of the {\sl HEGRA} telescopes, whose camera was upgraded with SiPM detectors.}) were based on 
vacuum photo-multipliers (PMTs): PMTs of the visible band (300 -- 600~nm) were used to register the main part 
of EAS Cherenkov radiation, while the PMTs of the middle-UV band (200 -- 300 nm) were used 
to register the UV emission of the EAS. 

\begin{figure*}
\centering
\includegraphics[height=12cm,width=12cm]{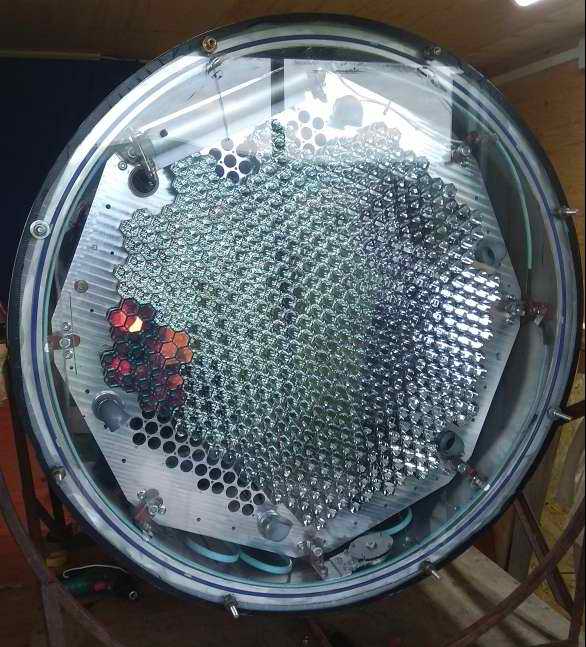}
\caption{
The front view (photosensitive side) of the TAIGA-IACT camera with an experimental 
SiPM cluster installed into camera in August 2022. 
Here the SiPM pixels of the experimental cluster (in the left side of camera) 
are covered by 290-590~nm band filters.
}
\label{fig2}
\end{figure*}

\begin{figure*}
\centering
\includegraphics[height=12cm,width=14cm]{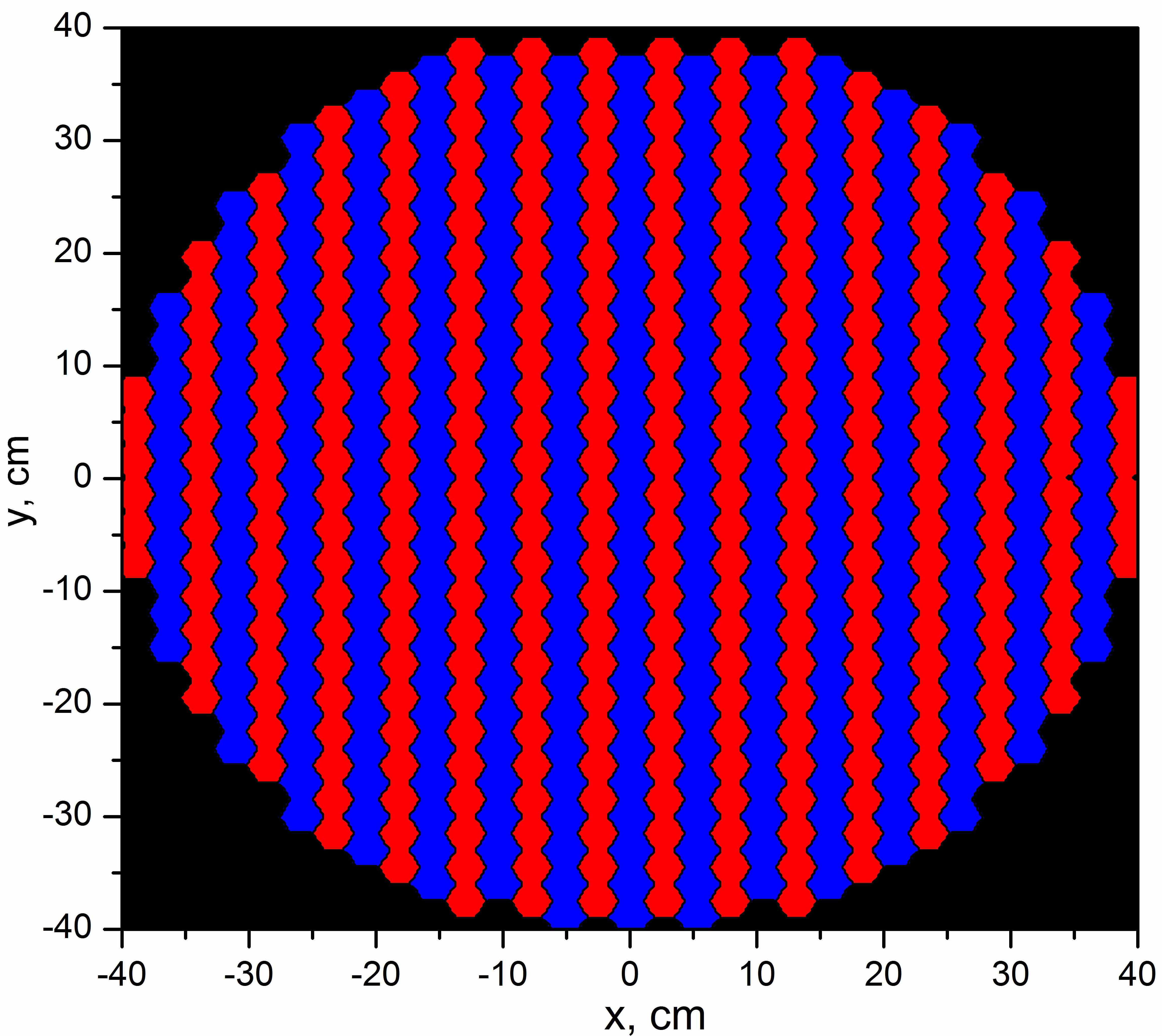}
\caption{An example of camera filling pattern. Two types of filters are employed: the pixels masked by (UVB+MUV) band (220-320~nm) filters 
are shown in red, the pixels masked by (visible+NUV) band (290-590~nm) filters are shown in blue.
}
\label{fig3}
\end{figure*}

One of the main changes being carried out in Cherenkov telescopes of the new (IV) generation, is the extensive 
use of silicon photo-multipliers (SiPMs) instead of vacuum PMTs as camera photo-detectors. After the successful 
operation of FACT (\citealt{Knoetig_et_al2013, Bretz_et_al2014}), several SiPM cameras were proposed and 
developed (see e.g. \citealt{Lombardi_et_al2020, Adams_et_al2021}). In accordance with this tendency, a new 
SiPM-based camera for small-size Cherenkov telescopes TAIGA-IACT is being developed at the Ioffe Institute  
(e.g. \citealt{Bogdanov_et_al2020b, Bogdanov_et_al2021a, Bogdanov_et_al2021b, Kuleshov_et_al2021, Bogdanov_et_al2022a}). 
TAIGA-IACT is an array of Cherenkov telescopes developed as an essential part of the gamma-ray and cosmic ray 
observatory TAIGA located in the Tunka valley, Rep. of Buryatiya. The TAIGA observatory consists of a set of 
instruments of various types, including wide-angle integrated timing arrays Tunka-133 and HiSCORE, 
scintillation arrays Tunka-Grande and TAIGA-muon arrays (e.g \citealt{Kuzmichev_et_al2018, Kuzmichev_2022, Budnev_et_al2022}). 
In August 2022 an experimental cluster based on OnSemi~MicroFJ-60035 SiPMs and 
optical filters SL 290 -- 590 manufactured by Photooptic Ltd\footnote{photooptic-filters.com}(\citealt{Photooptic_2020}) 
has been installed into the camera of one of the TAIGA-IACT units (see Fig. \ref{fig2}) and underwent full-scale operation test. 
One of the features of OnSemi~MicroFJ-60035 is a relatively wide-range~PDE (at a 6~V overvoltage the 
PDE$\ge 10\%$ in the 275-700~nm interval (\citealt{ON_Semi_Data_Sheet})). 
From one side, this requires usage of optical filters (at least to avoid the influence of the 
optical noise sharply increasing at wavelengths above $\simeq$640~nm due to the emission of 
atmospheric hydroxyl (OH) in the Meinel bands (e.g. \citealt{Leinert_et_al1998, Benn_Ellison1998}, see Fig. \ref{fig1}). 
But from the other side, this feature 
opens a possibility to use the same photosensors for registration of the EAS Cherenkov radiation 
in the middle UV band\footnote{The product of transmission of a (MUV+UVB)-band filter 
and the PDE of the chosen SiPM provides sensitivity in the 280-320\;nm band.} 
and in the visible band by application of filters with different transmission bands 
(see Fig. \ref{fig1}). 
The pixels of a telescope camera can be masked with two types of filters placed in 
some uniform order keeping the ``50\%/50\%'' number ratio  (e.g., Fig. \ref{fig3}). 
This would allow one to register simultaneously both the visible and 
the middle UV Cherenkov radiation from a particular flash (induced by a particular EAS) 
with one camera on a single standalone Cherenkov telescope, and thus, to perform gamma-hadron separation via a cut-off 
on the fraction of the middle UV band emission in the total flux of the measured Cherenkov flash emission. 
For suggested method of gamma-hadron separation the properties of (MUV+UVB)-filters are critically important while the properties of filters of main working range (visible + NUV) can 
vary in a wider range without significant worsening of efficiency of particle type discrimination. The key (MUV+UVB)-filter properties affecting gamma-hadron separation efficiency 
are the maximal transmission value and the transmission band $[\lambda_{1}, \lambda_{2}]$ (usually defined at a half of maximal transmission). The transmission values within this band should be 
as high as possible to provide the maximal level of signal, because the energy threshold to apply UV gamma-hadron separation is approximately iversely proportional to registration efficiency in (MUV+UVB)-range (ceteris paribus). 
The upper (longwavelength) limit $\lambda_{2}$ of transmission band should be in the interval of about $\sim$300 -- 320\;nm, where, from one side, the EAS Cherenkov spectrum is already sensitive 
to the primary particle type (see Fig. \ref{fig1}), and, from the other side, the number of EAS Cherenkov photons is still enough to be registered successfully. The lower (shortwavelength) limit $\lambda_{1}$ 
of transmission band should provide a large enough width $\left(\lambda_{2}-\lambda_{1}\right)\gtrsim 40$\;nm of transmission band to register the maximal number of (MUV+UVB) EAS Cherenkov photons (ceteris paribus). 
An attempt to use filters with transmission band in other parts of UV-range will make UV gamma-hadron separation quite ineffective: in UVA-range the spectrum shape is much less sensitive to the particle type (see Fig. \ref{fig1}) than needed, 
while in the most part of UVC-range the number of EAS Cherenkov photons becomes too small to be registered efficiently (see Fig. \ref{fig1}). 

The present paper is devoted to investigation of efficiency of a TAIGA-IACT unit
with an upgraded camera based on OnSemi~MicroFJ-60035~SiPMs in regards of gamma-hadron separation 
carried out by measurements of the fraction of meduim UV emission in the total amount 
of EAS Cherenkov emission. 
Previously, such an efficiency was estimated only in some general terms (\citealt{Rahman_et_al2001, 
Rahman_et_al2002, Kholupenko_et_al2018, Kholupenko_et_al2022}) and the calculated quantities were not the ones 
that can be directly measured. For example, in \cite{Razdan_et_al2002} the main characteristics of the 
gamma-hadron separation efficiency (the quality factor) remained indeterminate. 
Specific conditions considered in \cite{Rahman_et_al2001, Rahman_et_al2002, Razdan_et_al2002, Kholupenko_et_al2018} 
also were different from conditions of the TAIGA observatory. 
In the present paper all the values are connected to measurable quantities,
and the overall separation efficiency is estimated in general terms (similar to \cite{Rahman_et_al2001, 
Rahman_et_al2002, Kholupenko_et_al2018}) and, finally, in the commonly accepted terms of the quality factor. 
The considered conditions are specific for TAIGA-IACT, however similar calculations can be carried out for any
of the currently operating or planned IACTs. The calculations are based on multi-particle Monte Carlo 
simulations described in Sec.~\ref{MC_Simulations_Initial_Conditions}.

\section{Monte Carlo Simulations of EAS Development and Emission}
\label{MC_Simulations_Initial_Conditions}

Formation and propagation of Cherenkov emission from extensive air showers (EASs) was modelled with 
the CORSIKA~7.7400 package (\citealt{Heck_et_al1998}), which is the most used tool for modeling of EASs, 
often considered a reference for other codes employed for this purpose.

The initial conditions (setup) for the modeling we have carried out, were the following. 
The primary particles were 10, 31.62, and 100~TeV gamma-quanta and 31.62, 100, and 316.2~TeV protons. 
The energies of protons differed by the factor of $10^{0.5}$ for the following reason: 
the Cherenkov emission images of events from CR protons are weaker (i.e., have less 
$Size$\footnote{In Cherenkov astronomy, the $Size$ of a triggered event is the sum of the time-integrated 
signal in all the camera pixels, which are above the trigger.}) by a factor
of $\sim 1.5 - \sim 3$ than those from gamma-quanta of the same energy. 
Thus, images of EAS events from 31.62~TeV protons have the same average $\overline{Size}$ (which is a measured 
parameter) as images of events from gamma-quanta with energies in the [10, 31.62]~TeV interval 
(the exact value depends of a number of other factors, for details see Sec. \ref{Results}). 
The angle of incidence (the zenith angle, $\Theta$) of the primary particles was fixed at $40^{\circ}$. 
The direction of propagation of the primary particles was from South-to-North 
(X-axis; these conditions roughly imitate gamma-ray emission from the Crab Nebula 
as seen at the TAIGA observatory at its culmination).  
The telescope positions were [-180, -150, -120, -90, 0, 90, 120, 150, 180] m along X-axis (with Y=0 m) and 
[0, 90, 120, 150, 180] m along Y-axis (with X=0 m). 
The magnetic field strength was fixed at (H,Z)=(18.5, 57.7) $\mu$T (e.g. \citealt{Thebault_et_al2015}). 
The observation altitude at TAIGA is 675~m a.s.l. (\citealt{Vasilyev_et_al2009}), the corresponding atmosphere depth is $\simeq$950 -- 960~g$\cdot$cm$^{-3}$. 
The atmosphere model was N3, winter middle-latitude (\citealt{Kneizys_et_al1996}). 
The considered wavelength range of the Cherenkov photons was 200-700~nm. 
The bunch size (CERSIZ parameter of CORSIKA) was 1 photon. 
A typical shape of EAS Cherenkov spectrum is shown in Fig.~\ref{fig1}. 

Monte Carlo simulations of the night sky background (NSB) were performed 
with the TAIGA~Soft package (\citealt{Kholupenko_et_al2022}). 
The functional dependence of a smoothed and averaged NSB spectrum $\bar I_{\lambda}$ on the wavelength 
$\lambda$ is based on the data from \cite{Leinert_et_al1998}. 
Additional spectral features taken into account in the NSB simulations are the atomic oxygen line  
at 557.7~nm and the doublet at \{630.2 nm and 636.4 nm\} (approximately 
simulated as a line with a significantly broadened base). These lines are the most bright emission 
features of the night atmosphere in the 200-700~nm band 
(\citealt{Benn_Ellison1998, Mikhalev_et_al2001, Mikhalev_Medvedeva2002}). 
The increasing trend in the $\gtrsim 640$~nm band imitates 
emission of the atmospheric hydroxyl (OH) in the Meinel bands (\citealt{Leinert_et_al1998, Benn_Ellison1998}). 
The total intensity is normalized to $I=\int I_{\lambda}d\lambda=3\cdot 10^{12}$ 
photon$\cdot$m$^{-2}$s$^{-1}$ster$^{-1}$ 
in the 300-600~nm band based on the data of the TAIGA-IACT team (\citealt{Kuzmichev_et_al2018}), that is in 
good accordance with the data of \cite{Mirzoyan_Lorenz1994} for a wide-angle 
detector (i.e. with the field of view $\gg 1^{\circ}$). 

The telescope unit parameters employed for the modeling were as follows: 
the field of view is 9.72$^{\circ}$; the mirror consists of 34 segments of 60~cm diameter each 
(\citealt{Lubsandorzhiev_et_al2017}), thus the total mirror area is $S_{m}\simeq 9.6$ m$^2$, 
the radius of the mirror mechanical construction is $R_{T}\simeq 2.15$~m; the focal length is 4.75~m; 
the mirror albedo (the reflectivity coefficient) is 0.8; the transmission coefficient of 
the cover glass is 0.9; the transmission curves of the optical filters of main working range 
SL 290 -- 590 (\citealt{Photooptic_2020}) and (MUV+UVB)-range NSH+KCSH (\citealt{Manomenova_et_al2005, Rudneva_et_al2018}) 
are as shown in Fig.~\ref{fig1}; 
the considered filter mask 
is shown in Fig.~\ref{fig3}; 
the Winston cone transmission is 0.7 
(this conservative estimate is based on the calculations by \cite{Antonov_et_al2024}); 
the photon detection efficiency $\mbox{PDE}\left(\lambda\right)$ of SiPM OnSemi 
MicroFJ-60035 \cite{ON_Semi_Data_Sheet} is presented in Fig. \ref{fig1}; 
the time frame of event registration is 15~ns.

Monte Carlo simulations of photon transport in the telescope and of event detection in its 
camera also were carried out by means of the TAIGA~Soft package. 
All the details of the step-by-step algorithm realized in TAIGA~Soft can be found in 
\cite{Kholupenko_et_al2022,Kholupenko_et_al2024}.

\section{Results and Discussion}
\label{Results}
Multiparticle Monte Carlo simulations of EAS emission as well as of 
the Cherenkov photon transport in the atmosphere
and in the telescope optical system, carried out according to the approach and algorithms 
described in \cite{Kholupenko_et_al2022, Kholupenko_et_al2020}, were ultimately used to calculate 
the following ratio: 
\begin{equation}
UV={\max\left\{I_{u}\right\} \over \max\left\{I_{v}\right\}},
\label{UV_def}
\end{equation}
where $I_{i}$ is the signal amplitude in pixel $i$ (measured in photo-electrons [ph.e.]), 
the index $u$ runs over the pixels masked by the (UVB+MUV) band filters (220-320~nm), 
while the index $v$ runs over the pixels masked by (visible+NUV) band filters (290-590~nm). 

In numerical modeling the $UV$-value is calculated as a function of primary particle energy and distance of registration (from EAS axis). 
In actual observations the particle energy is not measured directly but estimated by means of $Size$ parameter which is total number of 
registered photoelectrons generated in camera by EAS Cherenkov radiation. The value of this parameter is approximately proportional to 
the particle energy but proportionality coefficient is different for primary gamma-quanta and CR protons and also depends on distance from 
the EAS axis to registartion point. The correspondence (as exact as possible taking into account statistical nature of parameters) can be 
stated as a result of numerical modeling for specific telescope (i.e. this correspodence is unique for every cherenkov telescope). 
Thus the calculation allowing comparison of values of $UV$ parameter for gamma-quanta and CR protons directly 
for the same values of $Size$, demands two stages: first, to determine the dependencies of mean $\overline{Size}$ (and its mean-square-deviation) 
on particle energy for CR protons and gamma-quanta, and, second, to simulate events from CR protons and gamma-quanta for the same values of $Size$ parameter 
for specific values of distance from EAS axis. Such a calculation would require very large computer resources (processor time, RAM, and disk space). 
But general statement 
(for all types of cherenkov telescopes) about $\overline{Size}$ relation is that for fixed value of particle energy $E$ the corresponding proton parameter, $Size_{p}$, and 
gamma-quantum parameter, $Size_{\gamma}$, satisfy the following inequality: 
\begin{equation}
\overline{Size}_{\gamma}\left(E/3\right)<\overline{Size}_{p}\left(E\right)<\overline{Size}_{\gamma}\left(E\right).
\label{Size_inequality}
\end{equation}

For considered configuration of TAIGA IACT (with the SiPM-camera and filters of 290-590~nm (visible+NUV) and 220-320~nm (UVB+MUV) bands), 
the validity of inequality (\ref{Size_inequality}) can be seen from comparison of data of Tab. \ref{Size_gamma_tab} and Tab. \ref{Size_protons_tab} 
under assumption of monotonic continuous (approximately proportional) dependence of $\overline{Size}$ from particle energy. 

\hspace*{-2.0cm}
{
\begin{table}[h]
\centering
\caption{Values of $\overline{Size}\pm \sigma_{Size}$ parameter [ph.e.] for $\gamma$ primaries}
\begin{tabular}{llll}
  \hline
Energy, TeV & 10 & 31.62 & 100  \\
   (x, m; y, m) &  &  &   \\
  \hline
  (-180; 0)  & 510 $\pm$ 73 & 1761 $\pm$ 218 & 6036 $\pm$ 653  \\
  \hline
  (-150; 0)  & 516 $\pm$  113 & 1869 $\pm$ 342 & 6671 $\pm$ 1081  \\
  \hline
  (-120; 0)  & 462 $\pm$  161 & 1813 $\pm$ 512 & 6906 $\pm$ 1692  \\
  \hline
  (-90; 0)  & 330 $\pm$ 188 & 1422 $\pm$ 641 & 5942 $\pm$ 2321  \\
  \hline
  (0; 0) & 563 $\pm$  267 & 2247 $\pm$ 1291 & 9351 $\pm$ 6134 \\
  \hline
  (90; 0)  & 299 $\pm$ 172 & 1294 $\pm$ 599 & 5449 $\pm$ 2224   \\
  \hline
  (120; 0) & 422 $\pm$ 153 & 1673 $\pm$ 486 & 6371 $\pm$ 1587  \\
  \hline
  (150; 0)  & 479 $\pm$ 107 & 1744 $\pm$ 320 & 6241 $\pm$ 995   \\
  \hline
  (180; 0) & 483 $\pm$ 70 & 1662 $\pm$ 201 & 5690 $\pm$ 592 \\    
  \hline
\\
  \hline
  (0; 0) & 563 $\pm$  267 & 2247 $\pm$ 1291 & 9351 $\pm$ 6134  \\
  \hline
  (0; 90)  & 612 $\pm$ 191 & 2319 $\pm$ 656 & 8795 $\pm$ 2334   \\
  \hline
  (0; 120)  & 667 $\pm$ 179 & 2488 $\pm$ 587 & 9139 $\pm$ 1831  \\
  \hline
  (0; 150)  & 716 $\pm$ 134 & 2557 $\pm$ 384 & 8883 $\pm$ 949 \\
  \hline
  (0; 180)  & 609  $\pm$ 47 & 2008 $\pm$ 115 & 6526 $\pm$ 216  \\
  \hline
\label{Size_gamma_tab}
\end{tabular}
\end{table}
}

\hspace*{-2.0cm}
{
\begin{table}[h]
\centering
\caption{Values of $\overline{Size}\pm \sigma_{Size}$ parameter [ph.e.] for proton primaries}
\begin{tabular}{llll}
  \hline
 Energy, TeV  &  31.62 & 100 & 316.2 \\
   (x, m; y, m) &  &  &  \\
  \hline
  (-180; 0)  & 973 $\pm$ 273 & 3739 $\pm$ 852 & 13541 $\pm$ 2611\\
  \hline
  (-150; 0)  & 1077 $\pm$ 389 & 4236 $\pm$ 1268 & 15830 $\pm$ 4095 \\
  \hline
  (-120; 0)  & 1137 $\pm$ 552 & 4642 $\pm$ 1992 & 17857 $\pm$ 6335\\
  \hline
  (-90; 0)  & 1112 $\pm$ 732 & 4687 $\pm$ 2835 & 18613 $\pm$ 8943 \\
  \hline
  (0; 0)  & 2476 $\pm$ 5697 & 11473 $\pm$ 19084 & 51920 $\pm$ 55127 \\
  \hline
  (90; 0)   & 1038 $\pm$ 678 & 4361 $\pm$ 2663 & 17295 $\pm$ 8386 \\
  \hline
  (120; 0)  & 1054 $\pm$ 514 & 4322 $\pm$ 1872 & 16625 $\pm$ 5945 \\
  \hline
  (150; 0)   & 1003 $\pm$ 363  & 3975 $\pm$ 1221 & 14805 $\pm$ 3844\\
  \hline
  (180; 0)  & 917 $\pm$ 258  & 3516 $\pm$ 797 & 12749 $\pm$ 2421 \\    
  \hline
\\
  \hline
  (0; 0)  & 2476 $\pm$ 5697 & 11473 $\pm$ 19084 & 51920 $\pm$ 55127 \\
  \hline
  (0; 90)  & 1426 $\pm$ 720 & 5865 $\pm$ 2621 & 22732 $\pm$ 8511 \\
  \hline
  (0; 120)  & 1364 $\pm$ 555 & 5486 $\pm$ 1816 & 20809 $\pm$ 5921 \\
  \hline
  (0; 150)  & 1267 $\pm$ 400 & 4944 $\pm$ 1188 & 18078 $\pm$ 3397 \\
  \hline
  (0; 180) & 1035 $\pm$ 252 & 3786 $\pm$ 655 & 13243 $\pm$ 1716 \\
  \hline
\label{Size_protons_tab}
\end{tabular}
\end{table}
}

The upper part of Tables \ref{Size_gamma_tab} and \ref{Size_protons_tab} (as well as other tables in the paper) 
describes changes of investigated quantities along axis X (South$\rightarrow$North), while the lower parts do changes 
of investigated quantities along axis Y (East$\rightarrow$West). The point (0 m, 0 m) is included in tables twice for 
convenience.

There are several reasons leading to formation of complex two-dimensional (surface) distribution of $\overline{Size}$-parameter 
in XY-plane, in comparison with the simple axis-symmetric case of vertical incedence of primary particle without magnetic field. 
One of these reasons is of artificial origin (i.e. connected with specific equipment construction), while the others come from EAS physical nature. 
These reasons are the following: 
\\1) Order of camera filling with filters is asymmetric relative to camera axes $x$ and $y$ (see Fig. \ref{fig3}): the filters of same type are placed 
along $y$-camera axis, while filters of different types alternate along $x$-camera axis. Thus the images of the same EAS observed with offset along 
X-axis and Y-axis include different fractions of pixels masked by visible and MUV filters. This leads to significant asymmetry of surface $\overline{Size}$-distribution 
relative to placement of telescope (observation point) on different axes: 
$\overline{Size}$-values observed with offset along Y are larger than ones observed with offset along X by $\sim$1.1 -- $\sim$2 times 
for gamma-quanta and by $\sim$1 -- $\sim$1.4 times for protons depending on particle energy and distance from EAS axis. 
This ratio is different for gamma-quanta and protons because proton images are more scattered than gamma-quanta ones. 
\\2) The presence of magnetic field of $\simeq 0.6$ Gs leads to deflection of trajectories of charged particles. 
Despite the fact that energy of considered primary particles are larger than 1 TeV, the most part of secondary electrons (and positrons) with energies above 
Cherenkov threshold (which are responsible for generation of Cherenkov photons) have energies less than $\sim$1 GeV (see e.g. \citealt{Giller_et_al2004}). 
Corresponding Larmor radii are less than $\sim$50 km and may be comparable with EAS length. This leads to significant asymmetry of angular distribution 
of EAS Cherenkov beam on azimuth angle (see e.g. \citealt{Homola_et_al2015}) and broadening of angular distribution in the direction perpendicular to geomagnetic 
field (see e.g. \citealt{Cummings_et_al2021}). This, in turn, leads to formation of ``elliptical'' profile of Cherenkov photon surface density\footnote{Lines of constant 
levels of  Cherenkov photon surface density are ellipses.}, broadened in the direction perpendicular to geomagnetc field (in our case, Y) with an additional contribution 
to eccentricity at the level of $\sim 0.1$, which affects surface $\overline{Size}$-distribution correspondingly. 
\\3) The observation plane (Earth surface) is not perpendicular to the EAS axis. This leads to formation of ``elliptical'' profile of Cherenkov photon surface density 
(in the same sense as in point 2)), broadened in the direction of inclination (in our case, X) with an additional cotribution to eccentricity at the level of $\simeq \sin \Theta \simeq 0.64$, 
which affects surface $\overline{Size}$-distribution correspondingly.
\\4) The inclination of EAS leads to different values of slant depth on Rayleigh scattering for Cherenkov photons coming to observation surface with positive and negative 
values of coordinate X. Corresponding contribution to relative difference between surface density (and $\overline{Size}$) in points $X$ and $-X$ can be estimated as 
$\delta^{S}_{R}\simeq -2\Lambda_{R}X\sin \Theta$ (the upper index ``S'' stands for $Size$), where $\Lambda_{R}$ is Rayleigh scattering coefficient near Earth surface, 
which takes values from $\simeq 10^{-5}$ m$^{-1}$ at 570 nm to $1.44\cdot 10^{-4}$ m$^{-1}$ at 300 nm (see e.g. \citealt{Bucholtz1995}). 
The relative difference, $\delta^{S}_{R}$, may achieve $\sim -5$\% for distant observation points $X=\pm 180$ m. 


The asymmetry of surface $\overline{Size}$-distribution, arising from cumulative action of factors 1 -- 4 listed above (together with other traditional difficulties of EAS physics), 
is difficult to be predicted precisely (at percent level) from general reasonings and simple analytical estimates. 

The inequality (\ref{Size_inequality}) allows us to avoid two-stage calculation and 
to estimate the efficiency of UV gamma-hadron separation by consequent comparison 
of $UV$-parameters for events from CR protons of energy $E_{p}$ with ones from gamma-quanta 
of energies $E_{\gamma}=E_{p}$ and $E_{\gamma}\simeq E_{p}/3$. 

The simplest way to see the effectiveness of UV gamma-hadron separation is to compare 
the mean values of $UV$ and its mean square deviations. These values at various distances 
from the EAS axis and for various energies of the primary particle are shown in 
Tab.~\ref{UV_gamma_tab} (for the gamma-quanta) and in Tab.~\ref{UV_protons_tab} 
(for the cosmic ray protons). 

\hspace*{-2.0cm}
{
\begin{table}[h]
\centering
\caption{Mean values and mean square deviations of the $UV$-parameter for gamma-quanta}
\begin{tabular}{lccc}
  \hline
  Energy, TeV & 10 & 31.62 & 100 \\
  (x, m; y, m) &  &  & \\
  \hline  
  (-180; 0) & $\left(3.25 \pm 1.56\right)\cdot 10^{-2}$ & $\left(4.07 \pm 1.46\right)\cdot 10^{-2}$ & $\left(4.76 \pm 1.1\right)\cdot 10^{-2}$ \\
  \hline
  (-150; 0) & $\left(2.31 \pm 1.29\right)\cdot 10^{-2}$ & $\left(2.78 \pm 1.21\right)\cdot 10^{-2}$ & $\left(3.4 \pm 1.06\right)\cdot 10^{-2}$ \\
  \hline
  (-120; 0) & $\left(2.16 \pm 0.95\right)\cdot 10^{-2}$ & $\left(1.95 \pm 0.77\right)\cdot 10^{-2}$ & $\left(2.3 \pm 0.82\right)\cdot 10^{-2}$ \\
  \hline
  (-90; 0) & $\left(4.77 \pm 2.66\right)\cdot 10^{-2}$ & $\left(3.83 \pm 1.53\right)\cdot 10^{-2}$ & $\left(2.77 \pm 0.74\right)\cdot 10^{-2}$ \\
  \hline
  (0; 0) & $\left(6.46\pm 7.78\right)\cdot 10^{-3}$ & $\left(5.93\pm 2.97\right)\cdot 10^{-3}$  & $\left(6.66 \pm 3.39\right)\cdot 10^{-3}$ \\
  \hline
  (90; 0) & $\left(4.47 \pm 2.98\right)\cdot 10^{-2}$ & $\left(4.35 \pm 1.76\right)\cdot 10^{-2}$ & $\left(3.04 \pm 0.86\right)\cdot 10^{-2}$  \\
  \hline
  (120; 0) &  $\left(2.30\pm 1.00\right)\cdot 10^{-2}$ & $\left(1.90\pm 0.70\right)\cdot 10^{-2}$ & $\left(2.1\pm 0.77\right)\cdot 10^{-2}$ \\
  \hline
  (150; 0) & $\left(2.07 \pm 1.18\right)\cdot 10^{-2}$ & $\left(2.51 \pm 1.19\right)\cdot 10^{-2}$ & $\left(3.08 \pm 1.05\right)\cdot 10^{-2}$ \\
  \hline
  (180; 0) & $\left(2.86\pm  1.51\right)\cdot 10^{-2}$ & $\left(3.62\pm 1.42\right)\cdot 10^{-2}$ & $\left(4.35 \pm 1.17\right)\cdot 10^{-2}$  \\    
  \hline
\\
  \hline
  (0; 0) & $\left(6.46\pm 7.78\right)\cdot 10^{-3}$ & $\left(5.93\pm 2.97\right)\cdot 10^{-3}$  & $\left(6.66 \pm 3.39\right)\cdot 10^{-3}$ \\
  \hline
  (0; 90) & $\left(8.14 \pm 9.84\right)\cdot 10^{-3}$ & $\left(5.55 \pm 1.77\right)\cdot 10^{-3}$ & $\left(4.7 \pm 0.9\right)\cdot 10^{-3}$  \\
  \hline
  (0; 120) & $\left(7.15 \pm 11.2\right)\cdot 10^{-3}$ & $\left(3.9 \pm 1.59\right)\cdot 10^{-3}$ & $\left(3.3 \pm 0.7\right)\cdot 10^{-3}$  \\
  \hline
  (0; 150) & $\left(8.24 \pm 14.2\right)\cdot 10^{-3}$ & $\left(3.71 \pm 2.56\right)\cdot 10^{-3}$ & $\left(2.83 \pm 0.84\right)\cdot 10^{-3}$ \\
  \hline
  (0; 180) & $\left(9.72 \pm 16.6\right)\cdot 10^{-3}$ & $\left(5.24 \pm 4.02\right)\cdot 10^{-3}$ & $\left(3.82 \pm 1.31\right)\cdot 10^{-3}$  \\
  \hline
\label{UV_gamma_tab}
\end{tabular}
\end{table}
}

\hspace*{-2.0cm}
{
\begin{table}[h]
\centering
\caption{Mean values and mean square deviations of the $UV$-parameter for the cosmic ray protons}
\begin{tabular}{lccc}
  \hline
  Energy, TeV & 31.62 & 100 & 316.2 \\
  (x, m; y, m) &  &  &  \\
  \hline
  (-180; 0)  & $\left(4.66 \pm 2.26\right)\cdot 10^{-2}$ & $\left(4.94 \pm 1.55\right)\cdot 10^{-2}$ & $\left(4.86 \pm 1.09\right)\cdot 10^{-2}$ \\
  \hline
  (-150; 0)  & $\left(3.99 \pm 1.98\right)\cdot 10^{-2}$ & $\left(4.08 \pm 1.52\right)\cdot 10^{-2}$ & $\left(4.01 \pm 1.05\right)\cdot 10^{-2}$ \\
  \hline
  (-120; 0)  & $\left(3.76 \pm 1.7\right)\cdot 10^{-2}$ & $\left(3.22 \pm 1.15\right)\cdot 10^{-2}$ & $\left(3.30 \pm 1.07\right)\cdot 10^{-2}$ \\
  \hline
  (-90; 0)  & $\left(4.96 \pm 2.37\right)\cdot 10^{-2}$ & $\left(4.09 \pm 1.3\right)\cdot 10^{-2}$ & $\left(3.55 \pm 0.93\right)\cdot 10^{-2}$ \\
  \hline
  (0; 0) & $\left(1.99 \pm 1.2\right)\cdot 10^{-2}$ & $\left(1.69 \pm 1.16\right)\cdot 10^{-2}$ & $\left(1.76 \pm 1.06\right)\cdot 10^{-2}$ \\
  \hline
  (90; 0)  & $\left(4.8 \pm 2.64\right)\cdot 10^{-2}$ & $\left(4.3 \pm 1.47\right)\cdot 10^{-2}$ & $\left(3.67 \pm 0.97\right)\cdot 10^{-2}$ \\
  \hline
  (120; 0) & $\left(3.72 \pm 1.62\right)\cdot 10^{-2}$ & $\left(3.12 \pm 1.11\right)\cdot 10^{-2}$ & $\left(3.14 \pm 1.03\right)\cdot 10^{-2}$ \\
  \hline
  (150; 0)  & $\left(3.67 \pm 1.95\right)\cdot 10^{-2}$ & $\left(3.79 \pm 1.51\right)\cdot 10^{-2}$ & $\left(3.81 \pm 1.09\right)\cdot 10^{-2}$ \\
  \hline
  (180; 0) & $\left(4.41\pm 2.19\right)\cdot 10^{-2}$ & $\left(4.64\pm 1.64\right)\cdot 10^{-2}$ & $\left(4.54\pm 1.08\right)\cdot 10^{-2}$ \\    
  \hline
\\
  \hline
  (0; 0) & $\left(19.9 \pm 12\right)\cdot 10^{-3}$ & $\left(16.9 \pm 11.6\right)\cdot 10^{-3}$ & $\left(17.6 \pm 10.6\right)\cdot 10^{-3}$ \\
  \hline
  (0; 90)  & $\left(16.1 \pm 6.5\right)\cdot 10^{-3}$ & $\left(10.8 \pm 3.1\right)\cdot 10^{-3}$ & $\left(8.7 \pm 1.82\right)\cdot 10^{-3}$ \\
  \hline
  (0; 120)  & $\left(13.3 \pm 6.4\right)\cdot 10^{-3}$ & $\left(7.85 \pm 2.79\right)\cdot 10^{-3}$ & $\left(5.98 \pm 1.25\right)\cdot 10^{-3}$ \\
  \hline
  (0; 150)  & $\left(10.9 \pm 5.8\right)\cdot 10^{-3}$ & $\left(6.78 \pm 2.98\right)\cdot 10^{-3}$ & $\left(5.33 \pm 1.84\right)\cdot 10^{-3}$ \\
  \hline
  (0; 180)  & $\left(12.6 \pm 8.1\right)\cdot 10^{-3}$ & $\left(7.87 \pm 3.72\right)\cdot 10^{-3}$ & $\left(6.46 \pm 2.53\right)\cdot 10^{-3}$ \\
  \hline
\label{UV_protons_tab}
\end{tabular}
\end{table}
}

One may assume naively that $\overline{UV}$ will not depend on telescope offsets $(X, Y)$ significantly, expecting that conditions of transfer for UV-photons 
will be close to ones for visible photons and, then, the ratio of photon numbers will be kept during propagation. But actually, there are also 
(as well as in case of $\overline{Size}$-distribution) several reasons leading to formation complex two-dimensional (surface) distribution of $\overline{UV}$-parameter in XY-plane. 
These reasons are similar to ones leading to surface $\overline{Size}$-distribution asymmetry but have additional features: 
\\1) Asymmetric order of camera filling with filters (see Fig. \ref{fig3}) and definition (\ref{UV_def}) of the $UV$-value lead to the following effect: 
a) in case of telescope shift along X-axis, the pixels hit by first and second maximal numbers of Cherenkov photons will be masked by filters of different types 
(due to alternation of filter types along $x$-camera axis) with high probability; b) in case of telescope shift along Y-axis, 
the pixels hit by first and second maximal numbers of Cherenkov photons will be masked by filters of the same type (due to the same filter types along $y$-camera axis) 
with high probability. This will lead to asymmetry of surface $\overline{UV}$-distribution relative to X and Y axes of about order of magnitude, that is much more expressed 
than for $\overline{Size}$. 
\\2) The strong dependence of Rayleigh scattering on photon wavelength ($\sim\lambda^{-4}$) leads to the following effect: the UV-photons are scattered much more intensively than 
visible photons (corresponding Rayleigh scattering coefficient in MUV range is from $\simeq 1.7\cdot 10^{-4}$ m$^{-1}$ at 290 nm to $\simeq 3.2\cdot 10^{-4}$ m$^{-1}$ at 250 nm, 
see e.g. \citealt{Bucholtz1995}). As a result of that, the ring-like region containing the most of generated Cherenkov photons (in our calculations $r=120,~150$ m) becomes relatively 
impoverished with UV-photons while in the outside region and internal circle the relative enrichment of UV-ratio is observed. 
\\3) As mentioned above, the inclination of EAS leads to different values of slant depth on Rayleigh scattering for Cherenkov photons coming to observation surface with positive and 
negative values of coordinate X. But due to strong dependence of Rayleigh scattering on photon wavelength this difference for (UVB+MUV) photons is larger on average than for 
the (visible+NUV) photons. Corresponding contribution to relative difference between $\overline{UV}$-values in points $X$ and $-X$ can be estimated as 
$\delta^{UV}_{R}\simeq -2\left(\overline{\Lambda}^{UV}_{R}-\overline{\Lambda}^{vis}_{R}\right)X\sin \Theta$, where $\overline{\Lambda}^{UV}_{R}$ and $\overline{\Lambda}^{vis}_{R}$ 
is Rayleigh scattering coefficients averaged over pixel acceptances with filters 220 -- 320 nm and 290 -- 590 nm respectively. The relative difference, $\delta^{UV}_{R}$, may achieve 
$\sim -8$\% for distant observation points $X=\pm 180$ m.
\\4) The inclination of EAS also leads to different values of slant depth on ozone absorption for Cherenkov photons coming to observation surface with positive and negative 
values of coordinate X. Corresponding contribution to relative difference between $\overline{UV}$-values in points $X$ and $-X$ can be estimated as 
$\delta^{UV}_{O_{3}}\simeq -2\Lambda_{O_{3}}X\sin \Theta$, where $\Lambda_{O_{3}}$ is ozone absorption coefficient near Earth surface, which can be estimated at a level values from 
$\sim 10^{-5}$ m$^{-1}$ at 300 nm up to $\sim 10^{-3}$ m$^{-1}$ at 260 nm to  (based on crossection given by \cite{Burrows_et_al1999, Orphal2003}). 
The relative difference, $\delta^{UV}_{O_{3}}$, may achieve $\sim -5$\% for distant observation points $x=\pm 180$ m. 

To see from the $\overline{UV}\pm\sigma_{UV}$ data how effective the gamma-hadron separation can be, one may use the following simple and natural measure of distribution separation 
(``distance'' between distributions):
\begin{equation}
\Delta^{i}_{k}={\left(\overline{UV}_{i}-\overline{UV}_{k}\right)\over \left(\sigma^2_{UV,i}+\sigma^2_{UV,k}\right)^{1/2}}
\label{Separation_measure}
\end{equation}
where indecies $i,~k$ stand for distribution parameters (particle type, particle energy and others). 
Results of calculations of values $\Delta^{i}_{k}$ are presented in Tab. \ref{Delta_tab}. 

\hspace*{-2.0cm}
{
\begin{table}[h]
\centering
\caption{Measure of separation of $UV$-distributions for CR protons and gamma-quanta}
\begin{tabular}{lcccc}
  \hline
  (x, m; y, m) & $\Delta^{p31}_{\gamma10}$ & $\Delta^{p31}_{\gamma31}$ & $\Delta^{p100}_{\gamma31}$ & $\Delta^{p100}_{\gamma100}$ \\
  \hline
 (-180; 0)  &     0.513 &   0.219 & 0.409 &   \red{0.095}  \\ 
  \hline
 (-150; 0)    &   0.711  &  0.521  &  0.669  &  0.367   \\ 
  \hline
 (-120; 0)   &    0.822  &  0.97  &  0.918  &  0.651  \\  
  \hline
 (-90; 0)    &   \red{0.053} &  0.401  &  \red{0.13}  &   0.882   \\ 
  \hline
 (0; 0)     &   0.94  &   1.13  &   0.916  &  0.847   \\ 
  \hline
 (90; 0)    &   \red{0.083} &  \red{0.142}  &  \red{-0.022} &  0.74  \\   
  \hline
 (120; 0)   &    0.746  &  1.03  &  0.93  &   0.755  \\  
  \hline
 (150; 0)   &    0.702  &  0.508  &  0.666 &   0.386  \\  
  \hline
 (180; 0)   &    0.583  &   0.303 &  0.47  &   \red{0.144}  \\  
  \hline
\\
  \hline
 (0; 0)   &    0.94   &  1.13  &   0.916  &  0.847  \\  
  \hline
 (0; 90)  &    0.675 &  1.57  &   1.47   &  1.89   \\  
  \hline
 (0; 120)  &   0.477  &  1.43  &   1.23  &   1.58  \\   
  \hline
 (0; 150)  &   \red{0.173}  &  1.13  &   0.781  &  1.28 \\    
  \hline
 (0; 180)  &   \red{0.156} &   0.814  &   0.48   &  1.03 \\    
  \hline
\label{Delta_tab}
\end{tabular}
\end{table}
}

\begin{figure*}
\centering
\includegraphics[height=10cm,width=14cm]{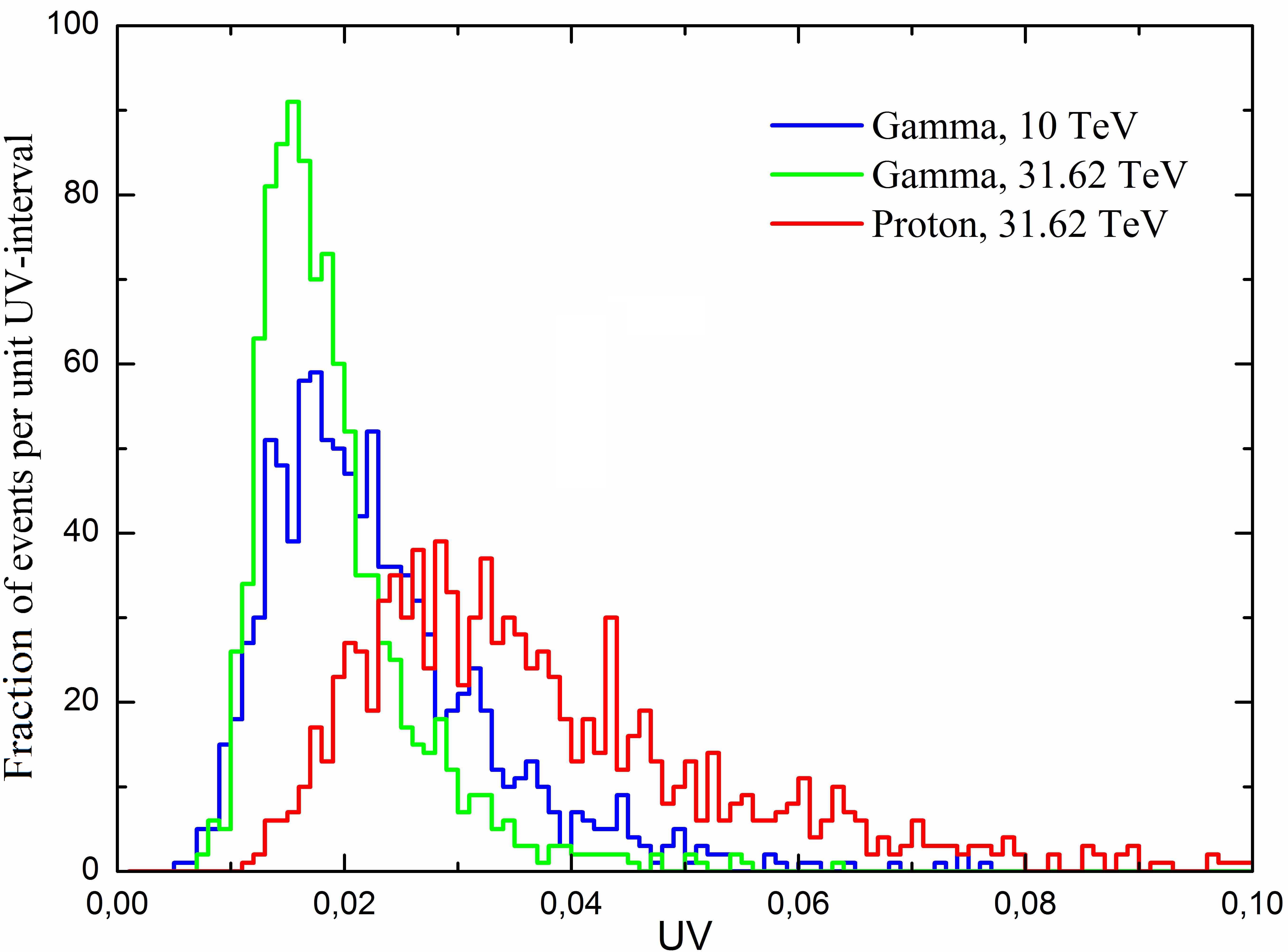}
\caption{
Distribution of the $UV$ parameter for events induced by 10~TeV (blue curve) and 31.62~TeV (green curve) gamma-quanta and by 
31.62 TeV protons (red curve) at observation point (120~m; 0 m). 
}
\label{fig4}
\end{figure*}

\begin{figure*}
\centering
\includegraphics[height=10cm,width=14cm]{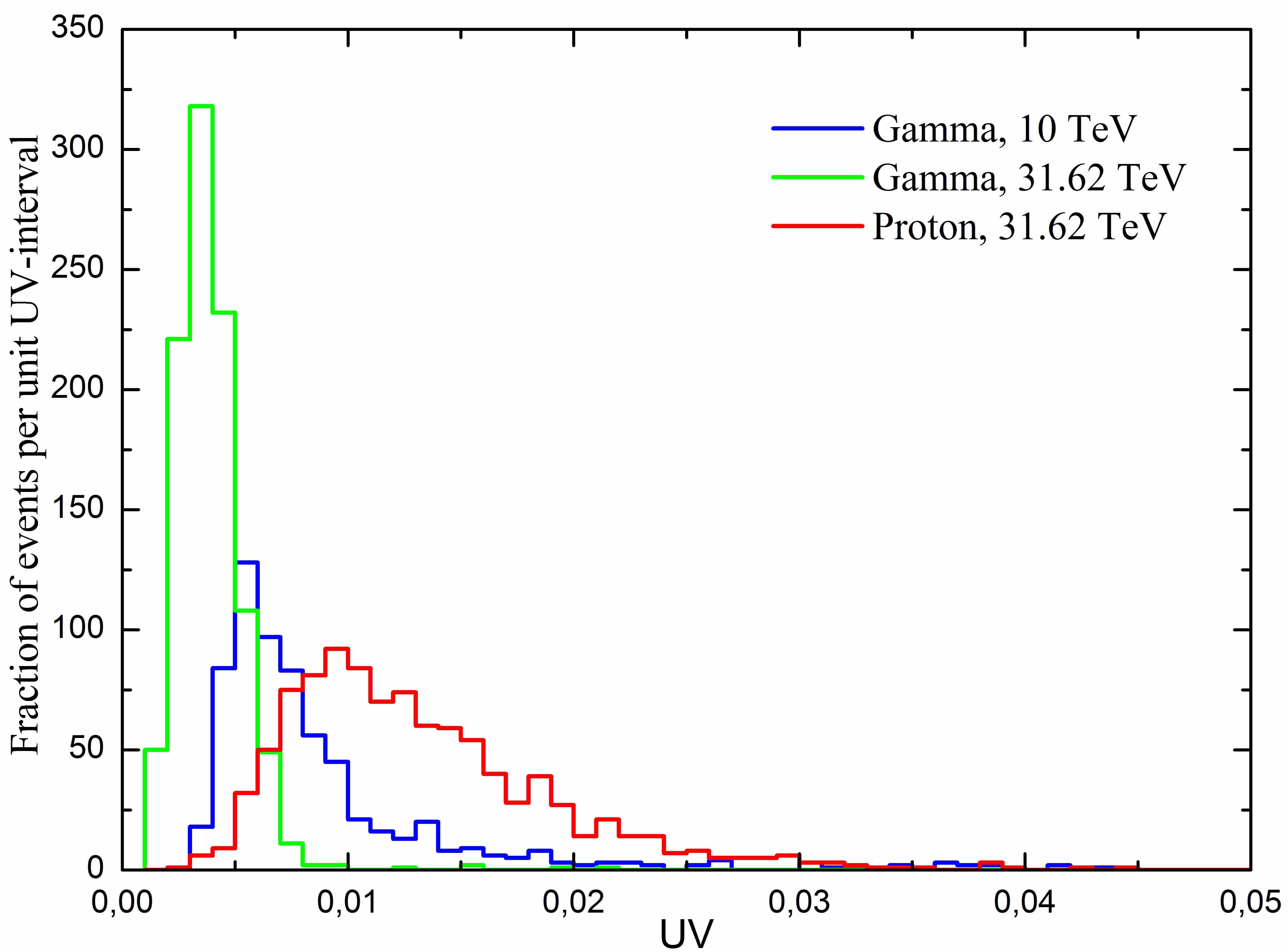}
\caption{
Distribution of the $UV$ parameter for events induced by 10~TeV (blue curve) and 31.62~TeV (green curve) gamma-quanta and by 
31.62 TeV protons (red curve) at observation point (0~m; 120~m). 
}
\label{fig5}
\end{figure*}

The general idea of using of measure (\ref{Separation_measure}) is the following: the larger the measure value the better 
the distribution separation and expected efficiency of gamma-hadron separation. Some deviations from this general tendence 
can be connected with the fact that actual $UV$-distributions are not Gaussian ones (and even not two-parametric ones). 
Despite these deviations one can make, for example, the following evident assumptions: if $\Delta^{i}_{k}\lesssim 0.2$ 
(these values are highlighted in Tab. \ref{Delta_tab} by red font) then 
the UV gamma-hadron separation will be almost impossible (since corresponding $UV$-distributions are overlapped strongly), 
if $\Delta^{i}_{k}\gtrsim 0.7$ then the UV gamma-hadron separation is expected to be effective enough (since corresponding 
$UV$-distribution are well-separated).
In accordance with these assumptions, the results presented in Tab. \ref{Delta_tab} show that UV gamma-hadron separation 
may be effective in the interval of distances $\sim [120; 150]$ m at telescope offset along X axis relative to EAS axis, and 
for all considered distances (excepting large distances for $\Delta^{p31}_{\gamma10}$) at telescope offset along Y axis relative to EAS axis. 
Examples of explicit view of separation of $UV$-distributions for 10 and 31.62~TeV $\gamma$-quanta and 31.62~TeV 
protons at observation points (120~m; 0 m) and (0 m; 120~m) are presented in Fig.~\ref{fig4} and \ref{fig5} respectively. 
The $UV$-distributions (the fractions of events per unit $UV$-interval, $f\left(UV\right)$) are normalized by the condition $\int f\left(UV\right)dUV=1$. 

Demanding only two parameters of $UV$-distribution, the calculation of separation measure (\ref{Separation_measure}) is the method 
for fast estimate of expected effectiveness of UV gamma-hadron separation for specific values of particle parameters. This method allows 
one to save the computer resources by rejecting exact calculations for obviously-ineffective variants of particle parameters\footnote{But 
in our simulations exact calculations for the full set of initial parameters are performed to state the correspondence between separation 
measure (\ref{Separation_measure}) and actual efficiency of UV gamma-hadron separation.}. 

Finally the efficiency of gamma-hadron separation is characterized by the quality factor $Q$. 
This factor shows the increase of the signal confidence level due to a particular selection criterion in 
comparison with the confidence level without selection (ceteris paribus) and is given by the following formula:
\begin{equation}
Q=\kappa_{\gamma}/\kappa_{p}^{1/2}\;, 
\label{Q_def}
\end{equation}
where $\kappa_{\gamma}$ and $\kappa_{p}$ are the acceptances of gamma- and proton-events, correspondingly 
(i.e. the fractions of selected events). 
These values can be calculated as follows:
\begin{equation}
\kappa_{i}\left(UV\le UV_{cr}\right)=\int_{0}^{UV_{cr}}f_{i}\left(UV\right)dUV, 
\label{acceptances}
\end{equation}
where subscript $i$ takes values ``$\gamma$'' and ``p'', $UV_{cr}$ is the $UV$-threshold defining a selection criterion. 

The Q-factors have been calculated by maximization of value of expression (\ref{Q_def}) at variation of $UV_{cr}$ with additional condition $\kappa_{p}\ge 0.025$. 
The latest is needed to provide stability of numerical results, 
because at small absolute values of $\kappa_{p}$ its relative numerical deviations may become very large, that, correspondingly, leads to large deviations of calculated 
$Q$-values. 
The results of calculations of acceptances $\kappa$ and corresponding maximal $Q$-factors, $Q_{max}$, for different values of particle energies 
and observation points are presented in Tab. \ref{Q_factor_g10_p31_tab} -- \ref{Q_factor_g100_p100_tab}. 
There are several typical cases: 
\\a) In many cases, the maximum is very flat: the relative width $\Delta UV_{cr}/UV^{max}_{cr}$ is much larger than 0.1 (typically $\sim 0.5$) at level of $0.95Q_{max}$. 
In this case the intervals of parameters, corresponding to $\sim [0.95Q_{max}; Q_{max}]$, are shown in the tables. The value of $UV_{cr}$ corresponding to $Q_{max}$ 
is approximately at the middle of interval shown in tables. Specific values of selection criterion $UV_{cr}$ can be chosen from additional reasonings, e.g. maximization of 
$\kappa_{\gamma}$ or minimization of $\kappa_{p}$ or other conditions arising from demands of Hillas analysis. 
\\b) In most cases, Q-factors achieve the maximal values at ``boundary'' condition $\kappa_{p}\ge 0.025$ (i.e. the next step to decrease $UV_{cr}$ leads to 
$\kappa_{p}< 0.025$). These values are marked by supperscript ``*'' in Tab. \ref{Q_factor_g10_p31_tab} -- \ref{Q_factor_g100_p100_tab}. 
\\c) In some cases, Q-factors achieve global maximumal values (including case when this is actually achieved at $\kappa_{p}=0.025$) and these maxima are expressive 
(i.e. sharp enough). These values are not marked by any additional indecies. 
\\d) In some cases, Q-factors remain below unity at all values of $UV_{cr}$. In these cases, values of selection criterion and acceptances are not presented (corresponding 
cells contain dashes ``--'').

\hspace*{-2.0cm}
{
\begin{table}[h]
\centering
\caption{Quality factor, Q, of separation between 10 TeV gamma-quanta and 31.62 TeV proton.}
\begin{tabular}{lcccccccc}
  \hline
   (x, m; y, m) & $UV_{cr}$ & $\kappa_{\gamma}$ & $\kappa_{p}$ & $Q$  \\
  \hline
  (-180; 0)  & 0.026 -- 0.067 & 0.409 -- 0.965 & 0.133 -- 0.756 & 1.11 -- 1.17  \\
  \hline
  (-150; 0)  & 0.014 -- 0.022 & 0.232 -- 0.565 & 0.025 -- 0.151 & 1.46 -- 1.54  \\
  \hline
  (-120; 0)  & 0.016 & 0.308 & 0.03 & 1.78$^{*}$  \\
  \hline
  (-90; 0)  & -- & -- & -- & $<$1  \\
  \hline
  (0; 0) & 0.007  & 0.515 & 0.028 & 3.08$^{*}$ \\
  \hline
  (90; 0)  & -- & -- & -- & $<$1  \\
  \hline
  (120; 0) & 0.017 & 0.298 & 0.032 & 1.67$^{*}$ \\
  \hline
  (150; 0)  & 0.014  & 0.308 & 0.032 & 1.72$^{*}$ \\
  \hline
  (180; 0) & 0.024 -- 0.048 & 0.461 -- 0.893 & 0.155 -- 0.594 & 1.16 -- 1.22 \\    
  \hline
 \\
  \hline
  (0; 0) & 0.007  & 0.515 & 0.028 & 3.08$^{*}$  \\
  \hline
  (0; 90)  & 0.007 & 0.257 & 0.027 & 1.56 \\
  \hline
  (0; 120)  & 0.006 & 0.232 & 0.048 & 1.06 \\
  \hline
  (0; 150)  & -- & -- & -- & $<$1  \\
  \hline
  (0; 180)  & -- & -- & -- & $<$1  \\
  \hline
\label{Q_factor_g10_p31_tab}
\end{tabular}
\end{table}
}

\hspace*{-2.0cm}
{
\begin{table}[h]
\centering
\caption{Values of quality factor, Q, of separation between 31.62 TeV gamma-quanta and 31.62 TeV proton.}
\begin{tabular}{lcccc}
  \hline
   (x, m; y, m) & $UV_{cr}$ & $\kappa_{\gamma}$ & $\kappa_{p}$  & $Q$ \\
  \hline
  (-180; 0)  &  0.054 -- 0.1 & 0.804 -- 1 & 0.599 -- 0.961 & 1.03 -- 1.08 \\
  \hline
  (-150; 0)  & 0.027 -- 0.05 & 0.587 -- 0.929 & 0.28 -- 0.711 & 1.1 -- 1.16 \\
  \hline
  (-120; 0)  & 0.016 & 0.388 & 0.03 & 2.24$^{*}$ \\
  \hline
  (-90; 0)  &  0.027 & 0.262 & 0.026 & 1.63$^{*}$ \\
  \hline
  (0; 0)  & 0.007 & 0.756 & 0.028 & 4.52$^{*}$\\
  \hline
  (90; 0)  & 0.037 -- 0.097 & 0.419 -- 0.993 &  0.142 -- 0.847 & 1.08 -- 1.14 \\
  \hline
  (120; 0)  & 0.017 & 0.478 & 0.032 & 2.67$^{*}$\\
  \hline
  (150; 0)  & 0.025 - 0.034 & 0.625 -- 0.822 & 0.285 -- 0.512 & 1.15 -- 1.21 \\
  \hline
  (180; 0) & 0.045 -- 0.088 & 0.762 -- 0.999 & 0.546 -- 0.945 & 1.03 -- 1.08 \\    
  \hline
 \\
  \hline
  (0; 0) & 0.007 & 0.756 & 0.028 & 4.52$^{*}$\\
  \hline
  (0; 90)  & 0.007 & 0.833 & 0.027 & 5.07$^{*}$\\
  \hline
  (0; 120) & 0.006 & 0.929 & 0.048 & 4.24$^{*}$\\
  \hline
  (0; 150)  &  0.005 & 0.871 & 0.059 & 3.59$^{*}$ \\
  \hline
  (0; 180)  & 0.004 & 0.415 & 0.025 & 2.63 \\
  \hline
\label{Q_factor_g31_p31_tab}
\end{tabular}
\end{table}
}

\hspace*{-2.0cm}
{
\begin{table}[h]
\centering
\caption{Values of quality factor, Q, of separation between 31.62 TeV gamma-quanta and 100 TeV proton.}
\begin{tabular}{lcccc}
  \hline
   (x, m; y, m) & $UV_{cr}$ & $\kappa_{\gamma}$ & $\kappa_{p}$  & $Q$ \\
  \hline
  (-180; 0)  & 0.045 -- 0.071 & 0.666 -- 0.957 & 0.438 -- 0.892 & 1.01 -- 1.054 \\
  \hline
  (-150; 0)  & 0.021 -- 0.027 & 0.334 -- 0.587 & 0.06 -- 0.188 & 1.36 -- 1.43 \\
  \hline
  (-120; 0)  & 0.017 & 0.462 & 0.028 & 2.76$^{*}$\\
  \hline
  (-90; 0)  & -- & -- & -- & $<$1 \\
  \hline
  (0; 0)  & 0.007 & 0.756 & 0.044 & 3.6$^{*}$ \\
  \hline
  (90; 0)  & -- & -- & -- & $<$1 \\
  \hline
  (120; 0)  & 0.018 & 0.548 & 0.044 & 2.61$^{*}$ \\
  \hline
  (150; 0)  & 0.018 & 0.304 & 0.038 & 1.56$^{*}$ \\
  \hline
  (180; 0) & 0.030 -- 0.050 & 0.411 -- 0.816 & 0.154 -- 0.632 & 1.03 -- 1.09 \\    
  \hline
 \\
  \hline
  (0; 0)  & 0.007 & 0.756 & 0.044 & 3.6$^{*}$ \\
  \hline
  (0; 90) & 0.006 & 0.666 & 0.034 & 3.61$^{*}$\\
  \hline
  (0; 120) & 0.004 & 0.589 & 0.032 & 3.29$^{*}$\\
  \hline
  (0; 150)  & 0.004 & 0.701 & 0.092 & 2.31$^{*}$ \\
  \hline
  (0; 180)  & 0.004 & 0.415 & 0.056 & 1.75$^{*}$ \\
  \hline
\label{Q_factor_g31_p100_tab}
\end{tabular}
\end{table}
}

\hspace*{-2.0cm}
{
\begin{table}[h]
\centering
\caption{Values of quality factor, Q, of separation between 100 TeV gamma-quanta and 100 TeV proton.}
\begin{tabular}{lcccc}
  \hline
   (x, m; y, m) & $UV_{cr}$ & $\kappa_{\gamma}$ & $\kappa_{p}$  & $Q$ \\
  \hline
  (-180; 0)  & 0.059 -- 0.090 & 0.829 -- 1 & 0.688 -- 1 &  1 -- 1.04\\
  \hline
  (-150; 0)  & 0.044 -- 0.074 & 0.814 -- 1 & 0.655 -- 0.98 &  1.01 -- 1.057 \\
  \hline
  (-120; 0)  & 0.017 & 0.255 & 0.028 & 1.52$^{*}$ \\
  \hline
  (-90; 0)  & 0.024 & 0.356 & 0.038 & 1.83$^{*}$ \\
  \hline
  (0; 0)  & 0.007 & 0.694 & 0.044 & 3.31$^{*}$\\
  \hline
  (90; 0)  & 0.024 & 0.247 & 0.03 & 1.43$^{*}$ \\
  \hline
  (120; 0) & 0.018 & 0.408 & 0.044 & 1.94$^{*}$\\
  \hline
  (150; 0) & 0.032 -- 0.1 & 0.64 -- 1 & 0.404 -- 1 & 1 -- 1.05 \\
  \hline
  (180; 0) & 0.060 -- 0.092 & 0.893 -- 1 & 0.79 -- 1 & 1 -- 1.04 \\    
  \hline
 \\
  \hline
  (0; 0) & 0.007 & 0.694 & 0.044 & 3.31$^{*}$\\
  \hline
  (0; 90) & 0.006 & 0.925 & 0.034 & 5.02$^{*}$ \\
  \hline
  (0; 120) & 0.004 & 0.875 & 0.032 & 4.89$^{*}$\\
  \hline
  (0; 150)  & 0.004 & 0.928 & 0.092 & 3.06$^{*}$ \\
  \hline
  (0; 180)  & 0.004 & 0.656 & 0.056 & 2.77$^{*}$ \\
  \hline
\label{Q_factor_g100_p100_tab}
\end{tabular}
\end{table}
}

The calculations show that measurements of the $UV$ parameter at an offset $|X|\simeq 120$~m from 
the EAS axis allows one to separate 31.62~TeV protons from 10~TeV and 31.62~TeV $\gamma$-quanta with 
a quality factor $Q\simeq 1.67 - 2.67$ under the criterion $UV\le 0.017$. 
For higher energies the calculations show that a measurement of the $UV$ parameter at the same offset 
allows one to separate 100~TeV protons from 31.62~TeV and 100~TeV $\gamma$-quanta with a quality factor 
$Q\simeq 1.52 - 2.76$ under the criterion $UV\le 0.017$. 
The values of Q-factor drop as distance to EAS axis increases (at distances larger than 120~m), but at offset $|X|\simeq 150$~m 
most of these values remain acceptable enough: $Q\simeq 1.05 - 1.72$ depending on particle energy. 
At offset $|X|\simeq 180$~m, the values of Q-factor drop below $\simeq 1.2$ and UV gamma-hadron separation becomes ineffective. 

For offsets along Y axis, the estimated Q-factors are significantly higher (expcepting $Q^{p31}_{\gamma 10}$ for $|Y|\ge 120$ m) and 
take values in the interval $1.56 - 5.07$. At that 12 of 18 estimated values are larger than 3, that says about very high potential efficiency 
of UV gamma-hadron separation for particles incident with offset along Y axis relative to the telescope position.

Here it should be emphasized once again that Q-factors are calculated as functions of telescope coordinates relative to the EAS axis. 
In general the Q-factors (for different types of gamma-hadron separation) can be calculated for specific values of distance (see e.g. \citealt{Razdan_et_al2002}) to EAS axis 
or for specific intervals of distance (i.e. averaged over the intervals, see e.g.\citealt{Chitnis_Bhat2001, Postnikov_et_al2017, Kunnas2017}). But Q-factor averaged over some interval 
of distance is naturally less than its peak values. Thus, if one can determine the distance with appropriate accuracy then the use of Q-factor as a function of distance 
(or averaged over narrow bins of distance) is preferable to achieve the more efficient separation where possible and to avoid the use of specific gamma-hadron separation 
where it is inefficient (i.e. at $Q<1$). Also, full (as possible) dependence of Q-factor on distance allows one to obtain the averaged Q-factors for distance intervals of any required size, 
while the inverse operation is impossible in general. 
Thus, the use of considered scenario of UV gamma-hadron separation demands determination of the EAS axis position as correct as 
possible\footnote{This is also usual demand for correct primary particle energy reconstruction (see e.g. \citealt{Ergin_phd2005})}. 
This, in turn, demands analysis of classical Hillas parameters (\citealt{Hillas1985}) for considered camera configuration (with SiPM-pixels masked by 
filters of two different types). Taking into account that the analysis of Hillas parameters can be performed at $Size \gtrsim$ 100 -- 120\;p.e. 
without doubts, the typical values of $Size$ estimated in the present paper for 10 TeV $\gamma$-EASs (see Tab. \ref{Size_gamma_tab}) and 31.62 TeV p-EASs  
(see Tab. \ref{Size_protons_tab}) allow one to assume that this analysis will be possible at TAIGA IACT with camera of considered configuration 
at energies of primary particles larger than 3 -- 4\;TeV. This, in turn, opens a possibility to use the stereoscopic mode (if there are two or more IACTs 
in the observatory, which is valid for TAIGA IACT array) to improve some characteristics of observatory (for example, angular resolution). 
The detailed investigations of efficiency of analysis of Hillas parameters and operation in stereoscopic mode for considered camera configuration are 
beyond the scope of the present paper and will be considered in the next papers. 


Obtained values of Q-factor are comparable with typical quality factors of gamma-hadron separation 
performed by analysis of single or pair of Hillas parameters ($length$ and $width$) for small-size Cherenkov telescopes 
(see, e.g., \citealt{Razdan_et_al2002, Postnikov_et_al2018}) and, in some cases, exceed them significantly. 
This makes UV gamma-hadron separation a perspective method to improve sensitivity of small-size Cherenkov telescopes. 
Here it is relevant to note that both Hillas and UV gamma-hadron separations are directed to the same aim (selection of gamma-quanta from whole set of observed particles), 
and both methods have significant estimated efficiency (if taking into account the results of the present paper on efficiency of UV gamma-hadron separation). 
This allows to assume the presence of some correlation between UV-parameter and Hillas parameters $length$ and $width$. The detailed estimate of this possible correlation 
is also beyond the scope of the present paper and will be investigated in the next papers. 

For convenient comparison of Q-factors with separation measure (\ref{Separation_measure}), the maximal calculated $Q$-values only have been 
collected together in the cumulative table \ref{Cumulative_Q_factor} (i.e. without intervals, additional indecies and comments). 

\hspace*{-2.0cm}
{
\begin{table}[h]
\centering
\caption{Cumulative table of Q-factors of UV gamma-hadron separation}
\begin{tabular}{lcccc}
  \hline
  (x, m; y, m) & $Q^{p31}_{\gamma10}$ & $Q^{p31}_{\gamma31}$ & $Q^{p100}_{\gamma31}$ & $Q^{p100}_{\gamma100}$ \\
  \hline
 (-180; 0)  &  1.17  & 1.08  &  1.054 &  1.04  \\ 
  \hline
 (-150; 0)    & 1.54  & 1.16  & 1.43 &  1.057  \\ 
  \hline
 (-120; 0)   & 1.78 & 2.24 & 2.76 & 1.52 \\  
  \hline
 (-90; 0)    &  \red{$<$1}  & 1.63 & \red{$<$1} & 1.83 \\ 
  \hline
 (0; 0)     &  3.08 & 4.52 & 3.6 & 3.31 \\ 
  \hline
 (90; 0)    &  \red{$<$1} & 1.14 & \red{$<$1} & 1.43 \\   
  \hline
 (120; 0)   & 1.67 & 2.67 & 2.61 & 1.94 \\  
  \hline
 (150; 0)   & 1.72 & 1.21 & 1.56 & 1.05 \\  
  \hline
 (180; 0)   & 1.22 & 1.08 & 1.09 & 1.04  \\  
  \hline
\\
  \hline
 (0; 0)   & 3.08 & 4.52 & 3.6 & 3.31 \\  
  \hline
 (0; 90)  &  1.56 & 5.07 & 3.61 & 5.02 \\  
  \hline
 (0; 120)  & 1.06 & 4.24 & 3.29 & 4.89 \\   
  \hline
 (0; 150)  & \red{$<$1} & 3.59 & 2.31 & 3.06 \\    
  \hline
 (0; 180)  & \red{$<$1} & 2.63 & 1.75 & 2.77 \\    
  \hline
\label{Cumulative_Q_factor}
\end{tabular}
\end{table}
} 

Comparison of data of Tab. \ref{Delta_tab} and Tab. \ref{Cumulative_Q_factor} shows that separation measure 
(\ref{Separation_measure}) allows one to predict expected level of Q-factor qualitatively. 
At $\Delta^{i}_{k}< 0.3$ all calculated values of Q-factor are less than 1.2, which means that UV gamma-hadron separation 
in these cases is ineffective. 
At $\Delta^{i}_{k}\ge 0.3$ one may note approximated relation $Q^{i}_{k}\simeq (2 - 4)\Delta^{i}_{k}$. 
This, for example means that at $\Delta^{i}_{k}=0.7$ the expected value of Q-factor can be estimated as $\gtrsim 1.4$, 
which means that UV gamma-hadron separation in these cases is effective enough. 

\section{Conclusion}

With multi-particle Monte Carlo simulations of performance of a small-size Cherenkov gamma-ray
telescope TAIGA-IACT equipped with modern SiPM detectors and two types of optical filters 
it is shown that the fraction of middle UV emission in the total amount of the detected 
Cherenkov signal can be efficiently used for gamma-hadron separation of the energetic primaries. 
In particular, measurements of only one $UV$-parameter at observation of Crab gamma-ray source 
would lead to primary type separation with quality factor $1.05\lesssim Q \lesssim 2.76$ at offsets 
120 -- 150 m along axis of camera filter alternation in the energy range 10 -- 100 TeV. 
Along perpendicular axis the quality factor is $1.75\lesssim Q \lesssim 5.07$ at offsets 0 -- 180 m 
in the energy range 31 -- 100 TeV and $1.06\lesssim Q \lesssim 5.07$ at offsets 0 -- 120 m 
in the energy range 10 -- 100 TeV. 
Hence, gamma-hadron separation based on the $UV$-parameter analysis can be considered as one of the perspective 
methods of detection quality improvement to be employed at future IACTs. 

In addition the dependencies of $Size$-parameter on particle energy for gamma-quanta and CR protons have been 
obtained specifically for the configuration of TAIGA-IACT equipped with SiPM detectors and two types of optical filters. 
These dependencies are neccessary for possibility of determination of particle energy from observable quantity $Size$, 
and will be useful for further simulations of EAS observations with this configuration of telescope.

\bigskip

{\bf Financial support} \\
The modeling by A.~B. was supported by grant of the Ministry of Science and Higher Education of the Russian Federation 23-075-67362-1-0409-000105. 
The work of A.~K. and E.~Kh. was supported by the baseline project FFUG-2024-0002 at the Ioffe Institute.
The work of N.~B. was supported by the baseline projects FZZE-2023-0004 and FZZE-2024-0005 at the Irkutsk State University. 

{\bf Conflict of interests} \\
The authors declare no conflicts of interest.

\bibliographystyle{raa}
\bibliography{ms2024-0211}
\end{document}